\documentclass[aps,prl,twocolumn,groupedaddress,floatfix]{revtex4-1}

\usepackage{graphicx}% Include figure files
\usepackage{epstopdf}
\usepackage{xcolor}

\usepackage{dcolumn} % Align table columns on decimal point
\usepackage{bm}      % bold math

\begin{document}

\title{Universal dynamics of a degenerate unitary Bose gas}

%% Notice placement of commas and superscripts and use of &
%% in the author list

\author{P. Makotyn, C. E. Klauss, D. L. Goldberger, E. A. Cornell \& D. S. Jin}
\affiliation{JILA, National Institute of Standards and Technology and University of Colorado, and Department of Physics, Boulder, CO 80309-0440, USA}

\begin{abstract}
From neutron stars to high-temperature superconductors, strongly interacting many-body systems at or near quantum degeneracy are a rich source of intriguing phenomena.  The microscopic structure of the first-discovered quantum fluid, superfluid liquid helium, is difficult to access due to limited experimental probes. While an ultracold atomic Bose gas with tunable interactions (characterized by its scattering length, $a$) had been proposed as an alternative strongly interacting Bose system ~\cite{ Cowell2002,Ho2004,Lee2010, Diederix2011, Li2012, Borzov2012,  jiang2013, Piatecki2013} , experimental progress ~\cite{papp2008,Navon2011,wild2012,smith2012} has been limited by its short lifetime.  Here we present time-resolved measurements of the momentum distribution of a Bose-condensed gas that is suddenly jumped to unitarity, \textit{i.e.} to $a=\infty$.  Contrary to expectation, we observe that the gas lives long enough to permit the momentum to evolve to a quasi-steady-state distribution, consistent with universality, while remaining degenerate. Investigations of the time evolution of this unitary Bose gas may lead to a deeper understanding of quantum many-body physics.
\end{abstract}

\maketitle

A powerful feature of atom gas experiments that provides access to these new regimes is the ability to change the interaction strength using a magnetic-field Feshbach resonance  ~\cite{Chin2010}. In particular, at the resonance location, $a$ is infinite. For atomic Fermi gases ~\cite{ohara2002,bartenstein2004,Regal2004,zwierlein2005a,partridge2006,stewart2006,giorgini2008}, accessing this regime by adiabatically changing $a$ led to the achievement of superfluids of paired fermions and enabled investigation of the crossover from superfluidity of weakly bound pairs, analogous to the Bardeen-Cooper-Schrieffer (BCS) theory of superconductors, to Bose-Einstein condensation (BEC) of tightly bound molecules~\cite{Regal2004, zwierlein2005a}. For bosonic atoms, however, this route to strong interactions is stymied by the fact that three-body inelastic collisions increase as $a$ to the fourth power~\cite{Shlyapnikov1996,Esry1999,Weber2003}.  This circumstance has limited experimental investigation of Bose gases with increasing interaction strength to studying either non-quantum-degenerate gases ~\cite{rem2013, fletcher2013} or BECs with modest interaction strengths ($na^3<0.008$, where $n$ is the atom number density)~\cite{papp2008,Navon2011,wild2012,smith2012}.

The problem is that the loss rate scales as $n^2 a^4$ while the equilibration rate scales as $n a^2 v$, where $v$ is the average velocity.  Thus, it would seem that the losses will always dominate as $a$ is increased to $\infty$.
Even if we were to forsake thermal equilibrium and suddenly change $a$ in order to project a weakly interacting BEC onto strong interactions~\cite{Claussen2002, Polkovnikov2011,smith2012, Sanner2012}, one might expect that three-body losses would still dominate the ensuing dynamics for large $a$.  In this work, however, we use this approach to take a BEC to the unitary gas regime, and we observe dynamics that in fact saturate on a timescale shorter than that set by three-body losses and that exhibit universal scaling with density.

One of the intriguing aspects of the unitary gas is that since $a$ diverges, it can no longer be a physically relevant scale for describing the system and its behavior.  For a gas near zero temperature, such as a BEC, the only physical scale that remains at unitarity is the interparticle spacing.  (In principle, the size of the cloud, or, equivalently the trap parameters, can provide a length scale, although one that is not intrinsic to the system. In addition, we are ignoring here any explicit three-body interactions, which could provide an additional length scale.) The gas behavior should then be universal in the sense that it is characterized only by the density $n$.  This means that energies scale as $n^{2/3}$, momenta as $n^{1/3}$, and times as $n^{-2/3}$, which we parameterize respectively  by  $E_n \equiv \hbar^2 (6 \pi^2 n)^{2/3}/2m$, $k_n \equiv (6\pi^2 n)^{1/3}$, and $t_n \equiv \hbar/E_n$.

The universality that makes the unitary gas so remarkable also provides a reason to hope that rapid three-body loss will not necessarily be an insurmountable barrier to experimental exploration of bulk (as opposed to lattice-confined) degenerate Bose gases with unitarity-limited interactions.  For the degenerate unitary Bose gas, both the loss rate and the equilibration rate must scale as $n^{2/3}$. The comparison of the two rates then hinges on unknown numerical prefactors, and it becomes an experimental question whether losses dominate or a \emph{local} equilibrium can be reached.
In addition, we note that on resonance, the shallow bound state that exists for finite positive $a$ disappears, so that loss requires atoms to decay to deeply bound molecular states~\cite{harter2013}. For $^{85} $Rb atoms, the previous experimental observation of a relatively narrow, and therefore long-lived, Efimov resonance (characterized by a dimensionless width, $\eta = 0.057 \ll 1$)~\cite{wild2012} is indicative that atoms close together do not decay instantaneously to deeply bound molecular states.

Our experiments (Fig.~1a) begin with a $^{85}$Rb BEC of between $5-7 \times 10^4$
atoms confined in a 10 Hz spherical magnetic  trap~\cite{pinothesis}.   The magnetic field, $B$, is set approximately 8 G above the $^{85}$Rb Feshbach resonance at $B_0=155.04$ G~\cite{Claussen2003}. This sets the initial $a$ to 142 $a_0$, which gives the BEC a Thomas-Fermi density distribution with an average density $\langle n \rangle=5.5(3) \times 10^{12}$ cm$^{-3}$.  With a typical initial temperature $< 10$ nK, the thermal deBroglie wavelength is large compared to $\langle n \rangle^{-1/3}$ and is not a relevant length scale in the physics of the ensuing experiment. Starting with this BEC in the extremely dilute limit, with $\langle n \rangle a^3 < 10^{-5}$,
we then decrease $B$ to $B_0$ in 5 $\mu $s. During the final 3 $\mu $s of the ramp of $B$, $\langle n \rangle a^3$ goes from an essentially dilute value of $10^{-4}$ to $\langle n \rangle a^3 \gg 1$.

After allowing the cloud to evolve at unitarity for a time $t$, we measure the momentum distribution of atoms by ramping, equally rapidly, back to small $a$ and allowing the gas to expand ballistically before imaging the cloud using resonant, high-intensity absorption imaging \cite{reinaudi2007}. From an azimuthal average of the image, we extract a momentum-space column density $\tilde{n}(\tilde{k})$ as a function of the component of momentum perpendicular to the line of sight, $\tilde{k}$.
By imaging at various times of flight (7, 13, 25 ms), we increase the dynamic range of our data and reduce the region of $\tilde{k}$ that is obscured by initial-size effects.
We repeat this experimental procedure for various $t$ to explore the evolution of the momentum distribution as a function of time at unitarity.

\begin{figure}
\includegraphics[width=8 cm]{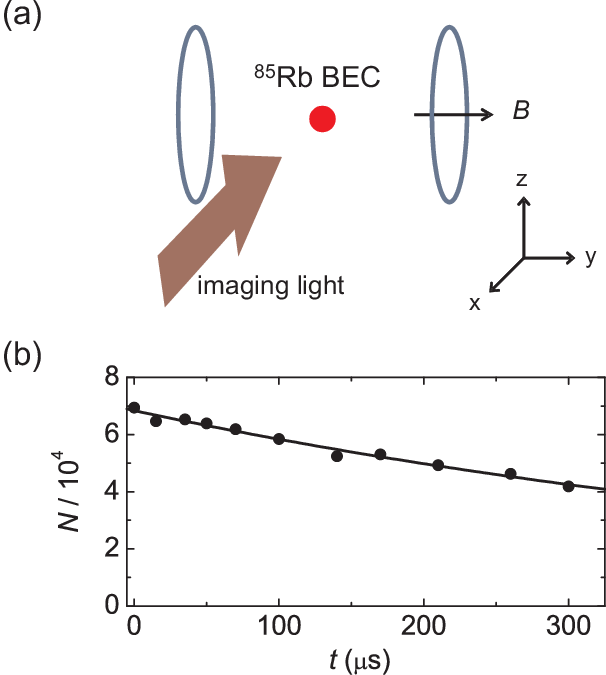}
\caption{a) Schematic showing the geometry of the imaging. The magnetic-field direction ($\hat{y}$), the imaging beam ($-\hat{x}$), and the direction of gravity (-$\hat{z}$) are mutually perpendicular. b) Number of atoms measured using absorption imaging as a function of the time at unitarity. The number measured without ramping to unitarity is shown at $t=0$. The solid line shows an exponential fit to the data (points), which gives a time constant of 630 $\pm$ 30 $\mu$s.
}
\label{fig:number}
\end{figure}

From images of the expanded cloud, we also obtain the number of atoms, $N$, which we show in Fig.~1b as a function of $t$. Fitting an exponential decay to this early time data yields a time constant of $630 \pm 30$ $\mu$s. In addition, the measured change in the spatial volume of the condensate is $(6\pm9) \%$ during the first 500 $\mu$s at unitarity.
A fact that is immediately clear from this data is that the density loss at unitarity occurs on a timescale that is much longer than the few $\mu$s duration of our ramps onto and away from the Feshbach resonance.  The ramp duration is also much shorter than the characteristic time set by the interparticle spacing, $t_n=57$ $\mu$s.

\begin{figure}
\includegraphics[width=7 cm]{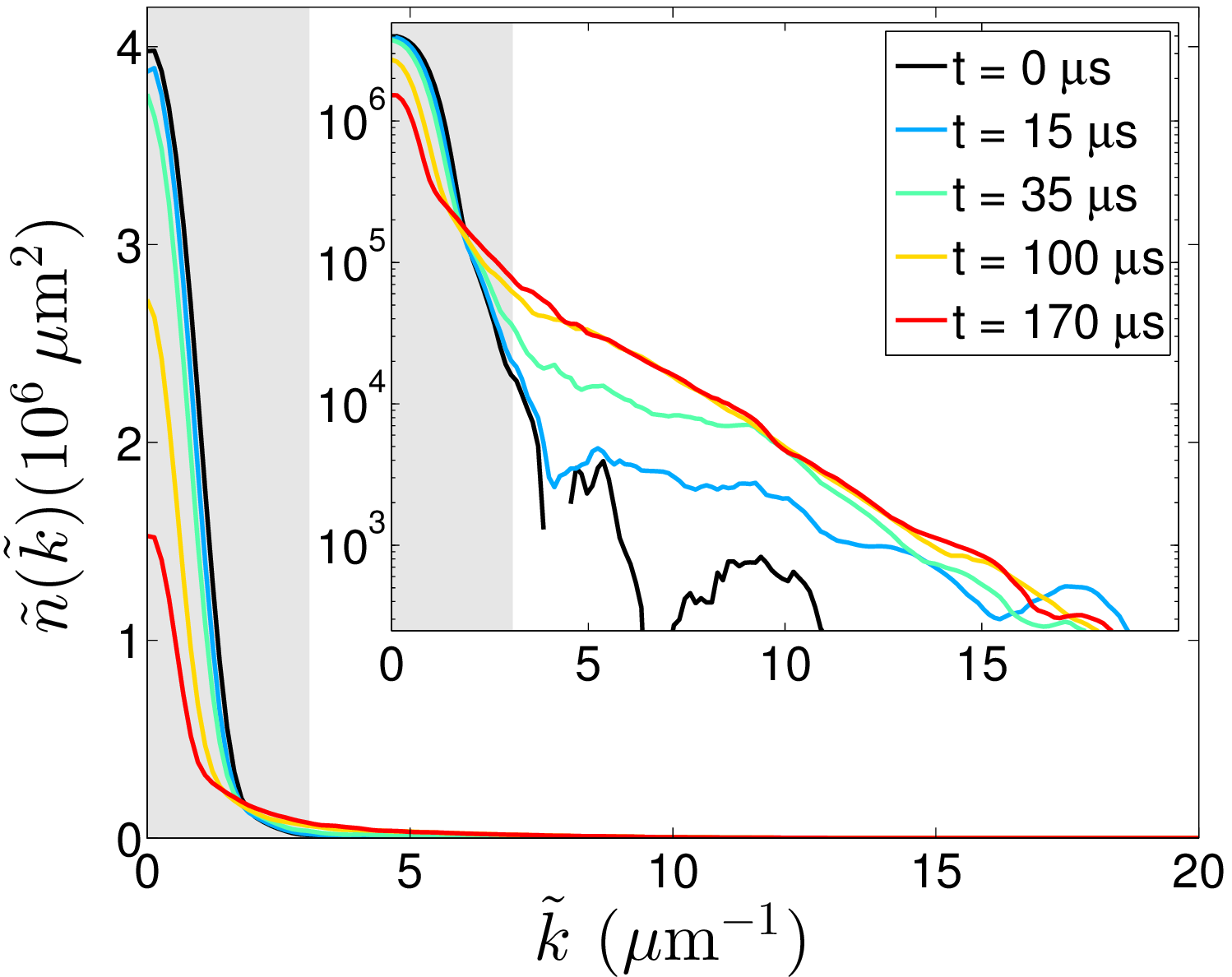}
\caption{The column-integrated momentum distribution $\tilde{n}(\tilde{k})$ versus the transverse momentum $\tilde{k}$ after evolving at unitarity for time $t$. The distribution measured without ramping to unitarity is shown at $t=0$. For each $t$, the integral $\int{\tilde{n}(\tilde{k})2\pi\tilde{k}{\rm d}\tilde{k}}=8\pi^3N(t)$. For this data $\langle n \rangle = 5.5(3) \times 10^{12} $ cm$^{-3}$, which corresponds to $k_n=6.9$ $\mu$m$^{-1}$ . Each momentum distribution is obtained from several images for each of three expansion times (7, 13, and 25 ms). The inset shows the same data plotted on log-linear axes. The gray regions indicate the part of data that is contaminated by initial-size effects and, therefore, does not accurately reflect the momentum distribution. We observe the emergence of signal outside this region, and a saturation of $\tilde{n}(\tilde{k})$ for $t>$100 $\mu$s.}
\label{fig:mom_dist}
\end{figure}

Equipped with this information regarding the timescales for number loss and for expansion of the trapped gas at unitarity,
we now consider the measured momentum distributions. These are shown in Fig. 2 for various $t$, with the inset showing the same data on a log-linear plot. Given the finite times of flight before imaging, the data at small $\tilde{k}$ are strongly affected by the initial size of the BEC and do not accurately reflect $\tilde{n}(\tilde{k})$; the gray regions in Fig. 2 indicate where initial-size effects are non-negligible, and we see that a significant fraction of the signal lies within this region. Nevertheless, the data clearly show the emergence of signal at high $\tilde{k}$, outside the gray regions. The signal at high $\tilde{k}$ grows as a function of $t$ before saturating in approximately 100 $\mu$s.  In this time, the gas has not yet lost a significant number of atoms or significantly reduced its density. The fact that the evolution timescale for $\tilde{n}(\tilde{k})$ is very different than the loss timescale clearly points to a mechanism for this dynamics that is distinct from three-body loss.  Furthermore, the much shorter timescale for saturation of $\tilde{n}(\tilde{k})$ suggests the existence of a  ``quasi-equilibrium'' metastable state of a degenerate Bose gas at unitarity.

\begin{figure}
\includegraphics[width=8 cm]{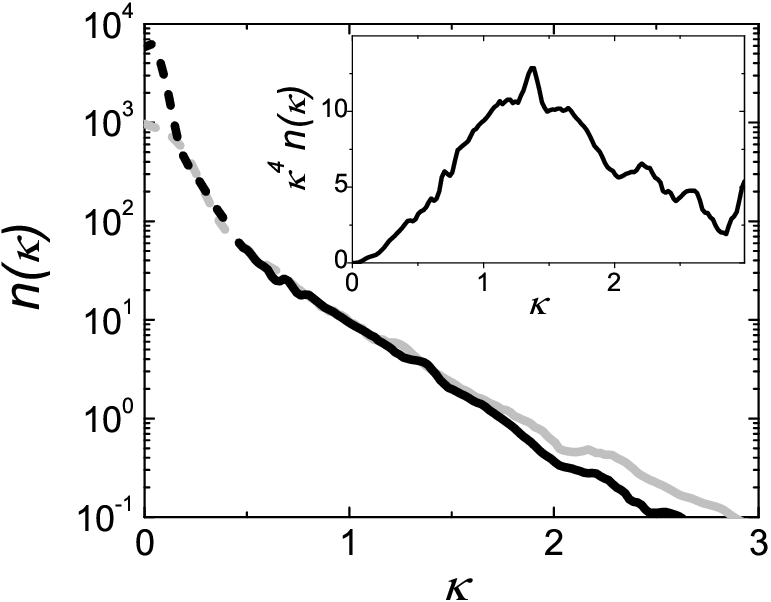}
\caption{The momentum distribution, $n(\kappa)$, plotted versus the scaled momentum, $\kappa$. Data for $\langle n \rangle=5.5(3) \times 10^{12}~$ cm$^{-3}$ and $\langle n \rangle=1.6(1) \times 10^{12}~$ cm$^{-3}$ are shown as the black and gray lines, respectively. Dotted lines indicate where the data are contaminated by finite-size effects. The higher $\langle n \rangle$ data is the average of measurements for 6 hold times $t$ between 100 $\mu$s and 300 $\mu$s, while the lower $\langle n \rangle$ data is the average of 4 measurements for $t$ between 200 $\mu$s and 700 $\mu$s. The distributions are normalized so that $\int{n(\kappa)4\pi \kappa^2 {\rm{d}} \kappa}=8 \pi^3$. The data for two different densities are consistent with a single curve when plotted in scaled units. Inset: Plotting $\kappa^4 n(\kappa)$ for high $\langle n \rangle$, we do not find clear evidence for a $1/\kappa^4$ tail at high $\kappa$.
}
\label{fig:collapse}
\end{figure}

To look for evidence of universality, we repeated the measurements for a lower initial density of the BEC. The measured $\tilde{n}(\tilde{k})$ for lower initial spatial density $\langle n \rangle$ also shows the emergence of signal at high $\tilde{k}$ at unitarity.  The distributions are similar to those measured for the higher $\langle n \rangle$ (Fig. 2), except that the dynamics occur over a longer time scale, with $\tilde{n}(\tilde{k})$ saturating in approximately 200 $\mu$s. To extract the three-dimensional $n(k)$, we use an inverse Abel transform. In  Fig. 3, we show the saturated momentum distributions as a function of the scaled momentum, $\kappa=k/k_n$, where $k_n$ is calculated at the average density $\langle n \rangle$. We find that the shape of the distributions for the two $\langle n \rangle$ are very similar.

Given that our data are consistent with a universal shape for the saturated $n(\kappa)$ at high $\kappa$, we now discuss aspects of this distribution.   First, we note that although much of the signal remains at small $\kappa$ where our data are affected by initial-size effects, the population with $\kappa>0.5$ for the saturated $n(\kappa)$ is nearly 50$\%$ of the initial $N$. Second, for two-body short-range interactions, such as those that give rise to the $s$-wave scattering length for atoms, one expects a $1/\kappa^4$  tail at high momentum for an equilibrium gas, where the amplitude of this tail is the thermodynamic parameter known as the contact~\cite{Tan2008}. We do not find evidence for a $1/\kappa^4$ tail at high momentum, which would appear as a flat line for large $\kappa$ in Fig. 2(inset); however, a $1/\kappa^4$ tail may exist below our detection limit at large $\kappa$ where the signal-to-noise ratio is poor.  In addition, three-body interactions could modify the high-momentum tail at unitarity \cite{Braaten2011}.

Finally, we consider the low-$\kappa$ part of the momentum distribution and the question of whether or not the gas remains degenerate after the rapid sweep to unitarity.
At low $\kappa$, initial-size effects can play a non-negligible role. However, this effect
is such that we can obtain a \textit{lower limit} on the fraction of atoms that have $\kappa<\kappa_{\mathrm{max}}$ by integrating our $n(\kappa)$ data up to $\kappa_{\mathrm{max}}$. This allows us to extract a lower limit for the density of atoms in phase space.  Specifically, we calculate the average occupancy per state at low $\kappa$, which is given by the number of atoms divided by the number of states in phase space:
\begin{equation}
\langle \rho_{occ} \rangle=\left(\frac{N}{8 \pi^3} \int\limits_0^{\kappa_{\mathrm{max}}}{n(\kappa) 4 \pi \kappa^2 \mathrm{d} \kappa}\right)/\left(\frac{V}{h^3} \frac{4 \pi}{3} (\hbar k_n)^3 \kappa_{\mathrm{max}}^3\right),
\end{equation}
where we conservatively use for the effective coordinate-space volume, $V=\frac{4 \pi}{3}R_{\mathrm{TF}}^3$, where $R_{\mathrm{TF}}$ is the Thomas-Fermi radius of the initial weakly interacting BEC.
For the higher $\langle n \rangle$ data, where the effects of the initial size are smaller, choosing for example $\kappa_{\mathrm{max}}=0.26$ gives 23\% of the atoms and $\langle \rho_{occ} \rangle=7.1$ for $t=170$ $\mu$s = 3 $t_n$.
The fact that this lower limit for the density in phase space is much larger than 1 for a significant fraction of the atoms indicates that the gas is degenerate.

In addition to considering the saturated $n(\kappa)$, we present the observed timescale for the dynamics in Figs. 4 and 5.
As can be seen in Fig. 2, the evolution of the momentum distribution is not uniform, with the higher momentum population saturating earlier. In Fig. 4, we plot the number of atoms, $\Delta N$, within a specific momentum range as a function of $t$, for two different ranges of momentum $\kappa$ and for two different initial densities $\langle n \rangle$.
In each case, we find that the number of atoms within the specific momentum range grows and then saturates. We find that the timescale for this saturation increases for smaller $\kappa$, and for smaller $\langle n \rangle$.  We fit the data for each different $\kappa$ range to an exponential and extract a time constant $\tau$.

\begin{figure}
\includegraphics[width=8 cm]{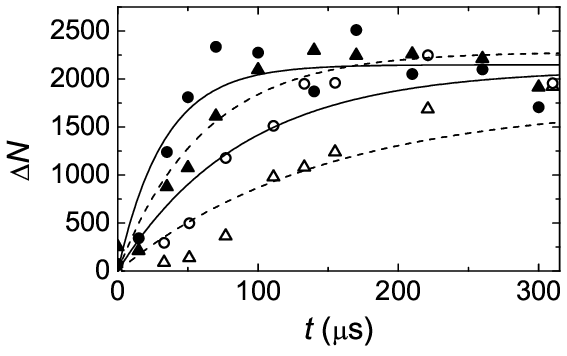}
\caption{The number of atoms in two momentum ranges vs. $t$. Data for $\langle n \rangle = 5.5(3) \times 10^{12} $ cm$^{-3}$ are shown with solid symbols, while data for $\langle n \rangle = 1.6(1) \times 10^{12} $ cm$^{-3}$ are shown with open symbols. The circles are the fraction of atoms with $\kappa$ between 1.20 and 1.32. The triangles are the fraction of atoms with $\kappa$ between 0.81 and 0.89. The lines show fits of the data to $\Delta N_0 (1-\exp^{-t/\tau})$, from which we extract the timescale for saturation, $\tau$.}
\label{fig:tau_fit}
\end{figure}

To look for universality in the timescales, we normalize $\tau$ by $t_n$, where $t_n$ is 57 $\mu$s and 130 $\mu$s for the data at higher and lower $\langle n \rangle$, respectively.
Plotting the normalized $\tau/t_n$ vs.  $\kappa$,  we find that the momentum-dependent dynamics at our two different densities are consistent (Fig. 5). We conclude that the timescale for $n(\kappa)$ dynamics is universal in that it depends only on the density, or interparticle spacing.  The momentum dependence of the timescales remains to be understood, although it is perhaps not unexpected that higher momenta dynamics saturate faster.

\begin{figure}
\includegraphics[width=8 cm]{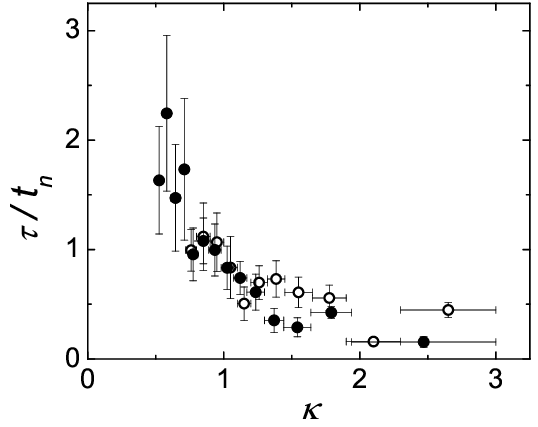}
\caption{The time constant associated with the emergence of signal at high momentum plotted as a function of scaled momentum, $\kappa$.  Data for $\langle n \rangle = 5.5(3) \times 10^{12} $ cm$^{-3}$ are shown with solid circles, while data for $\langle n \rangle = 1.6(1) \times 10^{12} $ cm$^{-3}$ are shown with open circles. Horizontal error bars show the spread in $\kappa$ used to extract each $\tau$, while the vertical bars correspond to 1 standard deviation.
}
\label{fig:tau_compiled}
\end{figure}

In conclusion, we have projected initially weakly interacting BECs onto unitarity-limited interactions and measured the resulting momentum-space dynamics. Three key findings of this work are as follows: (1) The momentum distribution of the unitary gas evolves and then saturates on a timescale that is significantly shorter that the timescale for three-body loss. (2) Both the shape of the saturated momentum distribution and the timescale for the dynamics appear to be universal. (3) The low-momentum part of the momentum distribution indicates that the density of atoms in phase space exceeds 1, and, hence, the gas is degenerate.  These findings support the conclusion that the gas reaches a locally equilibrated, metastable state and open the door for experimental investigation of a degenerate unitary Bose gas -- something that was previously considered inaccessible.

This work raises some interesting questions:
To what extent can the gas locally be described by a temperature, and is this temperature below the critical temperature for a Bose gas with unitarity-limited short-range interactions? What is this critical temperature in units of the critical temperature for an ideal Bose gas? At high momentum, what is expected for the contact, and does a high-momentum tail whose amplitude corresponds to the contact exist beyond the range of our data, or below our detection limit?  Finally, what does the observed momentum dependence of the dynamics tell us about the evolution of the system at unitarity?

\section*{Methods}
\textit{Magnetic-field control:} To rapidly change the magnetic field, we use an additional pair of coils, each with 10 turns and a diameter of 1.0 cm, 2.8 cm apart.
The step response of the system has a 10--$90 \%$ rise time of 2.1 $\mu$s; thus, the 5 $\mu$s magnetic-field sweep used in the measurements is below the maximum bandwidth of the system. We characterize and pre-correct for induced currents from mutual inductances between these coils and the magnetic trap coils as well as eddy currents in surrounding conductors.  Taking into account roughly equal contributions from uncertainty in our magnetic field and the uncertainty in the Feshbach resonance location $B_0$ \cite{Claussen2003}, we estimate that our experiments are within $\pm$ 50 mG of the Feshbach resonance, which corresponds to $|a|>95,000$ $a_0$.

\textit{Loss rate at unitarity:} Using the initial loss rate implied by the exponential fit to the data shown in Fig. 1, and using $\mathrm{d}N/\mathrm{d}t=-L_3 \int{ n(\mathbf{r})^3\mathrm{d}^3\mathbf{r}}$, we extract $L_3=5(1) \times 10^{-23}$ cm$^6$/s.  Unitarity-limited three-body loss rates for a non-degenerate Bose gas have been recently investigated by Rem \textit{et al.} \cite{rem2013}.
Using Eqn. 5 from Rem \textit{et al.} \cite{rem2013} and the Efimov resonance width, $\eta$, from Wild \textit{et al.} \cite{wild2012}, the predicted $L_3$ for $^{85}$Rb atoms at a temperature of 10 nK is $3 \times 10^{-20}$ cm$^6$/s, which is two and a half orders of magnitude larger than what we measure. On the other hand, after the jump to unitarity,
universality suggests that we should use an energy scale that is determined by the interparticle spacing. Replacing $k_B T$ with $E_n$, where $k_B$ is the Boltzmann constant, gives an estimate for $L_3$ of $1.7 \times 10^{-22}$ cm$^6$/s, which is within a factor of 4 of our measurement.
For the low $\langle n \rangle$ data, $L_3$ is a factor of 6.2(5) larger than for the high $\langle n \rangle$ data.  Using $E_n$ for the two different densities, we would expect this ratio to be 5.2(6).

\textit{Sample volume:}
The spherical aspect ratio of the trap was chosen to maximize the time before the cloud radius changes significantly.  {\it A priori}, it is not obvious whether the BEC will expand or collapse after the jump to unitarity. With in-situ images of the gas at unitarity, we find the cloud volume remains unchanged to within our measurement precision for $\sim$500 $\mu$s and then slowly increases. For experiments which require a lower initial coordinate space density, we begin the experimental cycle by changing $a$ to 400 $a_0$ for a fraction of a trap cycle, which transiently increases the cloud size and reduces the spatial density $\langle n \rangle$ to $1.6(1) \times 10^{12}~$ cm$^{-3}$.

\textit{Momentum distributions:} For the time-of-flight expansion, the 10 Hz spherical magnetic trap is turned off over 2 ms, while keeping the magnitude of the total magnetic field constant. Because the trap turns off in a time that is much shorter than the trap period, it has a negligible effect on the momenta of the atoms.  We image the atoms using a 5 $\mu$s imaging pulse.  The direction of the imaging beam and the magnetic-field direction are shown in Fig.~1b.  For each hold time at unitarity, we repeat the experiment four times for each of three different times of flight, $t_{\mathrm{exp}}$: 25 ms, 13 ms, and 7 ms.  Each image is azimuthally averaged, and the curves for the same TOF are averaged together.  We then combine the averaged curves into a single momentum distribution, $\tilde{n}(\tilde{k})$, using the largest $t_{\mathrm{exp}}$ data at the smallest $\tilde{k}$ and the smallest $t_{\mathrm{exp}}$ data at the largest $\tilde{k}$. This minimizes the initial-size effects at small $\tilde{k}$, while improving the signal-to-noise ratio at larger $\tilde{k}$.  In combining the curves, we enforce agreement in the overlap regions by applying a multiplicative factor to the data for shorter $t_{\mathrm{exp}}$.  This additional scaling factor, which ranges from 1.07 to 1.26 for the $t_{\mathrm{exp}}$=13 ms data and from 1.5 to 2.1 for the $t_{\mathrm{exp}}$=7 ms data, reflects systematic uncertainties that become increasingly important as $\tilde{n}(\tilde{k})$ decreases by orders of magnitudes.

At small $\tilde{k}$, the measured $\tilde{n}(\tilde{k})$ is distorted by the initial size of the BEC (the Thomas-Fermi radius is 16  $\mu$m for the higher $\langle n \rangle$ data and 22 $\mu$m for the lower $\langle n \rangle$ data) and by our imaging resolution (characterized by a gaussian width of approximately 6 $\mu$m).
The gray regions in Fig. 2, and the corresponding regions where the data are shown as dashed lines in Fig. 3, are bounded by a radius of 58 $\mu$m in the expanded cloud image.  In the absence of the jump to unitarity, the BEC with $\langle n \rangle = 5.5(3) \times 10^{12} $ cm$^{-3}$ has 97\% of the atoms within this radius after an expansion time of 25 ms. We note that all the effects discussed here cause low-momentum atoms to appear at larger radii than one would expect from the product of velocity and $t_{\mathrm{exp}}$.  Therefore, integrating the signal up to a particular momentum gives a lower limit to the number of atoms that have momenta below that value. We use this fact in extracting a lower bound for the density in phase space.

This material is based upon work supported by the National Science Foundation under Grant Number 1125844, the ONR, and the NIST.

The authors declare that they have no competing financial interests.
 
 Correspondence and requests for materials should be addressed to E.A.C. or D.S.J.

All the authors contributed to the experimental research described in the paper and to the writing of the manuscript.

%% Here is the endmatter stuff: Supplementary Info, etc.
%% Use \item's to separate, default label is "Acknowledgements"

\end{document}